\documentclass[aps,prd,twocolumn,nofootinbib,showpacs]{revtex4}

\usepackage{graphicx,color}
\usepackage[colorlinks=true, linkcolor=blue, citecolor=blue, urlcolor=blue]{hyperref}

\begin{document}

\title{Revisiting the Bottom Quark Forward-Backward Asymmetry $A_{\rm {FB}}$ in Electron-Positron Collisions}

\author{Sheng-Quan Wang$^1$}
\email[email:]{sqwang@cqu.edu.cn}

\author{Rui-Qing Meng$^1$}
\email[email:]{Rqing1008@163.com}

\author{Xing-Gang Wu$^2$}
\email[email:]{wuxg@cqu.edu.cn}

\author{Long Chen$^3$}
\email[email:]{longchen@mpp.mpg.de}

\author{Jian-Ming Shen$^4$}
\email[email:]{cqusjm@cqu.edu.cn}

\address{$^1$Department of Physics, Guizhou Minzu University, Guiyang 550025, P.R. China}
\address{$^2$Department of Physics, Chongqing University, Chongqing 401331, P.R. China}
\address{$^3$Max Planck Institute for Physics, F\"ohringer Ring 6, 80805 M\"unchen, Germany}
\address{$^4$School of Physics and Electronics, Hunan University, Changsha 410082, P.R. China}

\date{\today}

\begin{abstract}

The bottom quark forward-backward asymmetry $A_{\rm{FB}}$ is a key observable in electron-positron collisions at the $Z^{0}$ peak.
In this paper, we employ the Principle of Maximum Conformality (PMC) to fix the $\alpha_s$-running behavior of the next-to-next-to-leading order QCD corrections to $A_{\rm{FB}}$. The resulting PMC scale for this $A_{\rm{FB}}$ is an order of magnitude smaller than the conventional choice $\mu_r=M_Z$. This scale has the physically reasonable behavior and reflects the virtuality of its QCD dynamics, which is independent to the choice of renormalization scale. Our analyses show that the effective momentum flow for the bottom quark forward-backward asymmetry should be $\mu_r\ll M_Z$ other than the conventionally suggested $\mu_r=M_Z$. Moreover, the convergence of perturbative QCD series for $A_{\rm{FB}}$ is greatly improved using the PMC. Our prediction for the bare bottom quark forward-backward asymmetry is refined to be $A^{0,b}_{\rm FB}=0.1004\pm0.0016$, which diminishes the well known tension between the experimental determination for this (pseudo) observable and the respective Standard Model fit to $2.1\sigma$.


\end{abstract}

\maketitle

\section{Introduction}
\label{sec:1}

The experiments in electron-positron collisions have collected a wealth of data for the electroweak observables at the $Z^{0}$ peak. The bottom quark forward-backward asymmetry $A_{\rm FB}$ is a key precision electroweak observable, providing a stringent constraint on the effective electroweak mixing angle $\sin^{2}\theta^{\rm eff}_{W}$ as well as the mass of Higgs boson. However, $A_{\rm FB}$ shows the largest discrepancy with $2.9$ standard deviations between the experimental measurements and the Standard Model (SM) predictions among the measured set of precision electroweak observables~\cite{ALEPH:2005ab, ALEPH:2010aa}.

It is clearly important to understand whether this $A_{\rm FB}$ deviation is due to underestimated experimental uncertainties or inadequate theoretical calculations, or even physics beyond the SM. On the side of perturbative QCD (pQCD) corrections, which is the focus of this paper, the bottom quark forward-backward asymmetry have been calculated in Refs.\cite{Jersak:1981sp, Arbuzov:1991pr, Djouadi:1994wt} to the next-to-leading order (NLO) in $\alpha_s$. The next-to-next-to-leading order (NNLO) QCD calculations in the limit of massless $b$ quark
have been carried out in Refs.\cite{Altarelli:1992fs, Ravindran:1998jw, Catani:1999nf, Weinzierl:2006yt}.
It was noted that in the limit of massless quark $Q$, the forward-backward asymmetry is not infrared (IR) safe regardless of whether it is defined by the direction of flight of the quark $Q$ or by the thrust direction~\cite{Catani:1999nf}. For top quark pair production in electron-positron collisions, $A_{\rm FB}^t$ was computed at NNLO for the quark axis definition in~\cite{Gao:2014eea, Chen:2016zbz} using the partial result of~\cite{Bernreuther:2006vp}.
The $b$ quark forward-backward asymmetry was computed for $m_b\neq 0$ at NNLO both for the quark axis and the thrust axis definition in Ref.\cite{Bernreuther:2016ccf}, for conventional scale settings. It was found that this result reduces the tension between the measurements and the SM predictions but only slightly to $2.6 \sigma$.

All of the QCD calculations mentioned above have been obtained based on conventional scale setting, i.e.,~one simply sets the renormalization scale as the center-of-mass energy $\mu_r=\sqrt{s}=M_Z$, and the scale uncertainty is determined by varying the scale within certain range, e.g.,~$\mu_r\in[M_Z/2,2M_Z]$.
However, this \textit{ad~hoc} assignment of the renormalization scale, although conventional, introduces an inherent renormalization scheme-and-scale dependence in fixed-order predictions in perturbative QCD. One could obtain improper results if one applies this conventional procedure to QED processes~\cite{Wu:2013ei}. The conventional scale setting procedure also leads to the renormalon $n!\beta^n_0\alpha^n_s$ terms~\cite{Beneke:1998ui} and thus a nonconvergent perturbative series. In fact, the renormalization scale and effective number of flavors $n_f$ at each perturbative order are in general distinct, reflecting the different virtuality of the QCD dynamics.

In contrast, the Principle of Maximum Conformality (PMC)~\cite{Brodsky:2011ta, Brodsky:2012rj, Brodsky:2011ig, Mojaza:2012mf, Brodsky:2013vpa}
provides a systematic way to eliminate renormalization scheme-and-scale ambiguities in pQCD. The PMC provides the underlying principle for the Brodsky-Lepage-Mackenzie (BLM) method~\cite{Brodsky:1982gc}. The PMC scales are determined by absorbing the $\{\beta_i\}$ terms that govern the behavior of the running coupling via the renormalization group equation. Since those PMC scales are independent to the choice of renormalization scale, the PMC solves the conventional renormalization scheme and scale ambiguity from a basic way that agrees with the renormalization group invariance~\cite{Brodsky:2012ms, Wu:2014iba, Wu:2019mky}.
After applying the PMC, there is residual scale dependence due to unknown perturbative terms, which is different from the conventional renormalization scale dependence and is highly suppressed~\cite{Wu:2019mky}. Since the divergent renormalon terms disappear with the PMC scale setting, the convergence of pQCD series can be greatly improved.

We have shown that the large discrepancies between the experimental measurements and the SM predictions at the NLO QCD accuracy for the $t\bar{t}$ forward-backward asymmetry at hadron collisions can be attributed to an improper choice of the renormalization scale. Since the top-pair forward-backward asymmetry is dominated by the NLO-term where the scale uncertainty is huge, simply setting $\mu_r=m_t$ using conventional scale setting will cause unreliable predictions~\cite{Wang:2015lna}. By using the PMC, a comprehensive and self-consistent analysis for both the $t\bar{t}$ production cross-section and the $t\bar{t}$ forward-backward asymmetry can be obtained~\cite{Brodsky:2012rj, Brodsky:2012sz, Brodsky:2012ik, Wang:2014sua}; especially, the large discrepancy of the $t\bar{t}$ forward-backward asymmetry between the SM estimate and the CDF and D0 measurements are greatly reduced~\cite{Brodsky:2012rj}. Finally, after having higher-order QCD corrections and the complete SM NLO corrections systematically accounted for in Ref.\cite{Czakon:2017lgo} a very good agreement was observed between the experimental measurements and the SM predictions.

It is interesting to assess how much of the currently observed large deviation in the $A_{\rm FB}$ in electron-positron collisions could be due to the improper choice of the renormalization scales adopted in the theoretical computations.
In this paper, using the PMC method we will make a detailed analysis for this observable $A_{\rm FB}$ in electron-positron collisions in this regard.
The remaining sections of this paper are organized as follows. In Sec.\ref{sec:2}, we describe various technical aspects of applying PMC scale setting to the calculation of the $b$ quark forward-backward asymmetry at NNLO in QCD. In Sec.\ref{sec:3}, we present the numerical results with discussions on the improvements from the PMC method. The paper is concluded with a summary in Sec.\ref{sec:4}.

\section{PMC scale-setting for $b$ quark forward-backward asymmetry}
\label{sec:2}

\subsection{PMC scale-setting for the $b$ quark pair production cross section in electron-positron collisions}

For our PMC scale-setting analysis for the $b$ quark forward-backward asymmetry to NNLO in QCD, we employ the computational set-up of Refs.\cite{Chen:2016zbz, Bernreuther:2016ccf} for massive $b$ quark pair production in electron-positron collisions.
The NNLO QCD corrections consist of three classes of contributions:
(i) the two-loop and 1-loop squared corrections to the two-parton final state ($b\bar{b}$);
(ii) the one-loop corrections to the three-parton final state ($b\bar{b}g$); (iii) the tree level processes  with four-parton final states ($b\bar{b}gg$, $b\bar{b}q\bar{q}$, $b\bar{b}b\bar{b}$).

The QCD corrections for the $b$ quark pair production cross section can be conveniently divided into the (hard) non-Coulomb and Coulomb part, i.e.,
\begin{widetext}
\begin{eqnarray}
\sigma &=&\sigma^{(0)}\left[1+c^{(1)}_h\,a_s(\mu_r) + \left(c^{(2)}_{h,in}(\mu_r)+c^{(2)}_{h,n_f}(\mu_r)\,n_f\right)\,a^2_s(\mu_r) \right. \nonumber\\
&&\left. + \left(\frac{\pi}{v}\right)\,c^{(1)}_{v}\,a_s(\mu_r) + \left(\frac{\pi}{v}\right)\,\left(c^{(2)}_{v,in}(\mu_r)+c^{(2)}_{v,n_f}(\mu_r)\,n_f\right)\,a^2_s(\mu_r)+ \left(\frac{\pi}{v}\right)^2\,c^{(2)}_{v^2}\,a^2_s(\mu_r) + {\cal O}(a^3_s)\right],
\end{eqnarray}
\end{widetext}
where $a_s(\mu_r)={\alpha_s(\mu_r)}/{\pi}$, $\mu_r$ is the renormalization scale and $\sigma^{(0)}$ denotes the Born level cross section.
The coefficients $c^{(1,2)}_h$ are for the non-Coulomb corrections, and the coefficients $c^{(1,2)}_v$ and $c^{(2)}_{v^2}$ are for the Coulomb corrections.
The NNLO correction coefficients can be further split into the $n_f$-dependent and $n_f$-independent part, where $n_f$ is the number of active quark flavours (it is related to the $\beta_0$ term via $\beta_0=11\,C_A/3-4/3\,T_R\,n_f$, where $C_A=3$, $T_R=1/2$).

The Coulomb correction is proportional to powers of $(\pi/v)^n$ and it plays an important role in the threshold region. The quark velocity is given by
\begin{eqnarray}
v=\sqrt{1-\frac{4\,m_b^2}{s}},
\end{eqnarray}
where $s$ is the squared $e^{+}~e^{-}$ center-of-mass energy and $m_b$ is the mass of the $b$ quark.

The PMC scales must be determined separately for the non-Coulomb and Coulomb corrections~\cite{Brodsky:1995ds, Brodsky:2012rj, Wang:2020ckr}. After applying the PMC procedure, we obtain
\begin{eqnarray}
\label{eq:XwithPMCformula}
\sigma &=& \sigma^{(0)}\left[1+c^{(1)}_h\,a_s(Q_h) + c^{(2)}_{h,\rm con}(\mu_r)\,a^2_s(Q_h) \right. \nonumber\\
&&\left. + \left(\frac{\pi}{v}\right)\,c^{(1)}_v\,a_s(Q_v) + \left(\frac{\pi}{v}\right)\,c^{(2)}_{v,\rm con}(\mu_r)\,a^2_s(Q_v)\right. \nonumber\\
&&\left. + \left(\frac{\pi}{v}\right)^2\,c^{(2)}_{v^2}\,a^2_s(Q_v) + {\cal O}(a^3_s)\right],
\end{eqnarray}
where, $Q_h$ and $Q_v$ are the PMC scales for the non-Coulomb and Coulomb corrections, respectively. $c^{(2)}_{h,\rm con}(\mu_r)$ and $c^{(2)}_{v,\rm con}(\mu_r)+\frac{\pi}{v}\,c^{(2)}_{v^2}$ are the corresponding conformal coefficients.

We found that the scale $Q_h$ associated with the (hard) non-Coulomb corrections is of the order $m_b$, since it originates from the hard gluon virtual corrections and is determined for the short distance process. As expected, the scale $Q_v$ for the Coulomb corrections is of the order $v\,m_b$, and depends continuously on the quark velocity $v$, as it originates from Coulomb rescattering. It is noted that the scale $Q_v$ becomes soft for $v\rightarrow0$. Thus, the PMC scales have the physically reasonable behavior and reflect the virtuality of the QCD dynamics~\cite{Wang:2020ckr}. Also the number of active flavors $n_f$ changes with the PMC scales.

\subsection{PMC scale-setting for the $b$ quark forward-backward asymmetry}

The $b$ quark forward-backward asymmetry $A_{\rm FB}$ is defined by
\begin{eqnarray}
A_{\rm FB}=\frac{N_{F}-N_{B}}{N_{F}+N_{B}},
\end{eqnarray}
where $N_F$ and $N_B$ are the number of the (massive) $b$ quarks observed in the forward and backward hemisphere, respectively.
The asymmetry $A_{\rm FB}$ can also be expressed in terms of the symmetric cross section $\sigma_S$ and the antisymmetric cross section $\sigma_A$ for the inclusive production of the $b$ quark, i.e.,
\begin{eqnarray}
A_{\rm FB}=\frac{\sigma_A}{\sigma_S}=\frac{\sigma_F-\sigma_B}{\sigma_F+\sigma_B}.
\end{eqnarray}
The $\sigma_F$ and $\sigma_B$ are the forward and backward cross sections, respectively, which can be written in terms of differential cross sections as
\begin{eqnarray}
\sigma_F&=&\int_0^1d\cos\theta \int_{x_0}^1dx\frac{d\sigma}{dx\,d\cos\theta}, \\
\sigma_B&=&\int_{-1}^0d\cos\theta \int_{x_0}^1dx\frac{d\sigma}{dx\,d\cos\theta},
\end{eqnarray}
where $\theta$ is the angle between the electron three-momentum and the axis defining the forward hemisphere. The energy ratio $x$ is defined as $2\,E_{b}/\sqrt{s}$ where $E_{b}$ is the energy of the $b$ quark. Apparently $x$ is bounded from below by $x_{0}=2\,m_{b}/\sqrt{s}$.

The QCD calculation for the forward-backward asymmetry $A_{\rm FB}$ to NNLO can be parameterized as
\begin{eqnarray}
A_{\rm FB}&=&A^{\rm LO}_{\rm FB}\,\left[1 + \delta^{(1)}\,a_s(\mu_r)+\left(\delta^{(2)}_{\rm in}(\mu_r) \right.\right. \nonumber\\
&& \left.\left. +\, \delta^{(2)}_{n_f}(\mu_r)\,n_f\right)\,a^2_s(\mu_r) + {\cal O}(\alpha_s^3)\right],
\end{eqnarray}
where $A^{\rm LO}_{\rm FB}$ denotes the forward-backward asymmetry at the Born level. The $\delta^{(1)}$ and $\delta^{(2)}$ are the coefficients of the NLO and NNLO QCD corrections respectively.
The NNLO coefficients $\delta^{(2)}(\mu_r)$ are given numerically in Ref.\cite{Bernreuther:2016ccf} both for the quark axis and thrust axis definition of the forward hemisphere.
In order to identify the $\beta_0$ term and then apply the PMC method properly, we computed the $b$ quark forward-backward asymmetry at NNLO both for the thrust axis and the quark axis definitions using the results of Ref.\cite{Bernreuther:2016ccf}.

The $b$ quark forward-backward asymmetry has been measured in electron-positron collisions at the $Z^0$ peak, where the center-of-mass energy is much larger than the $b$ quark mass.
Consequently, the quark velocity is $v\sim1$, and the Coulomb correction is negligible.
We thus only need to determine the PMC scale for the hard virtual corrections.
Since only the $\beta_0$-term appears at the present NNLO level, we have just one PMC scale for the forward-backward asymmetry $A_{\rm FB}$. We found that there is very small scale dependence from the NNLO-term for the asymmetry $A_{\rm FB}$. The PMC single-scale approach (PMC-s) provides a rigorous way to eliminate the scale dependence from the last term~\cite{Shen:2017pdu}. The PMC-s approach fixes the renormalization scale by directly requiring all the RG-dependent nonconformal terms up to a given order to vanish; thus it inherits most of the features of the PMC standard multiscale approach. Its predictions are also scheme independent due to the resulting conformal series~\cite{Wu:2018cmb}. Thus, in order to eliminate the scale dependence from the last NNLO-term and ensure the renormalization scheme independence, we adopt the PMC-s approach to do the scale-setting for the asymmetry $A_{\rm FB}$, we then obtain
\begin{eqnarray}
A_{\rm FB}&=&A^{\rm LO}_{\rm FB}\,\left[1 + \delta^{(1)}\,a_s(\mu^{\rm PMC}_r) \right. \nonumber\\
&&\left. +\, \delta^{(2)}_{\rm con}(\mu_r)\,a^2_s(\mu^{\rm PMC}_r ) + {\cal O}(\alpha_s^3)\right].
\label{AFBPMC}
\end{eqnarray}
The PMC scale is determined as
\begin{eqnarray}
\mu^{\rm PMC}_r=\mu_r\,\exp\left[\frac{3\,\delta^{(2)}_{n_f}(\mu_r)}{2\,T_R\,\delta^{(1)}}\right],
\end{eqnarray}
and the conformal coefficient can be written as
\begin{eqnarray}
\delta^{(2)}_{\rm con}(\mu_r)&=&\delta^{(2)}_{\rm in}(\mu_r)+\frac{11\,C_A}{4\,T_R}\,\delta^{(2)}_{n_f}(\mu_r)\,.
\end{eqnarray}

At the present NNLO level, the PMC scale $\mu^{\rm PMC}_r$ and conformal coefficient $\delta^{(2)}_{\rm con}(\mu_r)$ are only formally depends on the choice of the renormalization scale $\mu_r$, their own values are independent of the choice of renormalization scale $\mu_r$~\cite{Wu:2013ei}. Thus, the resulting PMC prediction in Eq.(\ref{AFBPMC}) eliminates the renormalization scale uncertainty. The scale-independent results will be given in detail below. Since the $\beta_0$-term is absorbed into the coupling constant via the renormalization group equation, the correct argument of the coupling constant is thus determined. The purpose of PMC is not to find an optimal renormalization scale but to find the effective coupling constant (whose argument is called as the PMC scale) with the help of renormalization group equation. Since the effective coupling is independent to the choice of the renormalization scale $\mu_r$, thus solving the conventional scale ambiguity.

\section{Numerical results and discussions}
\label{sec:3}

For our numerical calculations, we use the RunDec program~\cite{Chetyrkin:2000yt} to evaluate the two-loop $\overline{\rm MS}$ scheme coupling from $\alpha_s(M_Z)=0.1181$. The input parameters are taken to be the same as those of Ref.\cite{Bernreuther:2016ccf}, e.g., we take the $b$-quark pole mass $m_b=4.89$ GeV, which is converted from the $\overline{\rm MS}$ mass $m_b^{\overline{\rm MS}}=4.18$ GeV, the sine of the weak mixing angle $\sin^2\theta_W=0.2229$, the electromagnetic coupling $\alpha=1/132.233$ and $G_\mu=1.166379\times10^5$ GeV$^{-2}$.

\subsection{The $b$ quark forward-backward asymmetry with thrust axis definition }

\begin{table}
\begin{tabular}{|c||c|c|c|c|c|c|}
\hline
~~ ~~  & ~~$\mu_r$~~ &~~LO~~  &~~NLO~~  &~~NNLO~~ &~~Total~~ &~~R~~ \\
\hline
~~ ~~ & $M_Z/2$ & 1 & -0.0321 & -0.0099 & 0.9580 & ~31\%~  \\
~Conv.~& $M_Z$ & 1 & -0.0287 & -0.0108 & 0.9605 & ~38\%~  \\
~~ ~~ & $2M_Z$ & 1 & -0.0260 & -0.0112 & 0.9628 & ~43\%~  \\
\hline
~PMC~ & & 1 & -0.0436 & -0.0035 & 0.9529 & ~8\%~ \\
\hline
\end{tabular}
\caption{The QCD correction factors $A_{\rm FB}/A^{\rm LO}_{\rm FB}$ for the $b$ quark forward-backward asymmetry with the thrust
axis definition at LO, NLO, NNLO using the conventional (Conv.) and PMC scale settings. The pure NNLO numbers for conventional scale settings agree with those given in Table 1 of Ref.~\cite{Bernreuther:2016ccf} at the percent level.
\label{tab1}  }
\end{table}

In Table \ref{tab1} we present the QCD correction factors $A_{\rm FB}/A^{\rm LO}_{\rm FB}$ for the $b$ quark forward-backward asymmetry with the thrust axis definition at LO, NLO, NNLO using the conventional and PMC scale settings. The LO term provides the dominant contributions and is free from QCD interactions. In the case of conventional scale setting, the total QCD correction factor is $0.9605$ for $\mu_r=M_Z$, and its scale uncertainty is only $0.5\%$ by varying $\mu_r\in[M_Z/2,2M_Z]$. Such a small scale uncertainty is due to cancelations among different contributing terms specific to $A_{\rm FB}$, and the scale uncertainty is rather large for each perturbative term. The NLO term is -0.0287 for $\mu_r=M_Z$ and its scale uncertainty is $21\%$ by varying $\mu_r\in[M_Z/2,2M_Z]$. The NNLO term is $-0.0108$ for $\mu_r=M_Z$ and its scale uncertainty is $12\%$ by varying $\mu_r\in[M_Z/2,2M_Z]$. Simply fixing the renormalization scale $\mu_r=M_Z$ might give a reasonable prediction for the total QCD correction; however, one cannot decide what is the exact QCD correction terms for each perturbative order.

We define a parameter
\begin{eqnarray}
R=(A_{\rm FB}/A^{\rm LO}_{\rm FB})|_{\rm NNLO}/(A_{\rm FB}/A^{\rm LO}_{\rm FB})|_{\rm NLO},
\end{eqnarray}
to show the relative importance of the NNLO-term and the NLO-term. Table \ref{tab1} shows that the values of $R$ are very large and they are changed from $31\%$ to $43\%$ for the scale $\mu_r\in[M_Z/2,2M_Z]$. Thus, the conventional predictions have a slow convergence in their perturbative series, which also depend on the choice of the scale $\mu_r$. After applying PMC scale setting, the renormalization scale uncertainty is eliminated. Moreover, due to the absorption of the divergent renormalon terms, the NLO correction term is largely increased while the NNLO correction term becomes very small compared to those in the conventional predictions. This ratio $R=(A_{\rm FB}/A^{\rm LO}_{\rm FB})|_{\rm NNLO}/(A_{\rm FB}/A^{\rm LO}_{\rm FB})|_{\rm NLO}$ in the case of PMC prediction is about $8\%$ for $\mu_r=M_Z$, and is not changed for a wide range of the scale $\mu_r$. Thus the convergence of the perturbative series is significantly improved after using the PMC.

\begin{figure}
\centering
\includegraphics[width=0.45\textwidth]{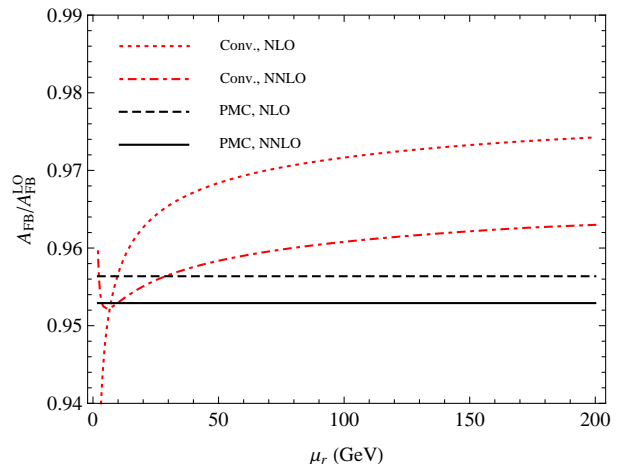}
\caption{The QCD correction factors $A_{\rm FB}/A^{\rm LO}_{\rm FB}$ versus the renormalization scale $\mu_r$ for the $b$ quark forward-backward asymmetry with the thrust axis definition using the conventional (Conv.) and PMC scale settings, where the dashed and solid lines stand for the PMC predictions up to the NLO and NNLO level, respectively; the dotted and dot-dashed lines are the corresponding results for conventional scale settings. The renormalization scale $\mu_r$ is changed in a wide range of $\mu_r\in$[2~GeV, 200~GeV].}
\label{convPMC1}
\end{figure}

In Fig.(\ref{convPMC1}) we plot the QCD correction factors $A_{\rm FB}/A^{\rm LO}_{\rm FB}$ versus the renormalization scale $\mu_r$ for the $b$ quark forward-backward asymmetry with the thrust axis definition using the conventional and PMC scale settings. Figure (\ref{convPMC1}) shows that using conventional scale setting, the QCD correction at NLO depends heavily on the choice of the scale $\mu_r$, whereas the scale dependence of the QCD correction at NNLO becomes smaller. This observation is consistent with the expectation that the scale dependence is progressively decreased by the inclusion of higher-order calculations. The total NNLO QCD correction factor (with the conventional scale setting) first decreases and then increases with increasing $\mu_r$, achieving its minimum value at $\mu_r\sim 6$ GeV. In contrast, after using the PMC, the scale dependences of both the separate QCD corrections at each order and the total QCD correction are simultaneously eliminated. The total QCD correction factor with the PMC setting is almost fixed to be $A_{\rm FB}/A^{\rm LO}_{\rm FB}=0.9529$ for a wide range of the scale $\mu_r$. This PMC-improved result is smaller compared to the conventional results obtained at $\mu_r=M_Z$ and is very close to the minimum of the conventional results obtained at $\mu_r\sim 6$ GeV.

More explicitly, we determine the PMC scale to be
\begin{eqnarray}
\mu^{\rm PMC}_r&=&9.7 ~\rm {GeV}.
\end{eqnarray}
for the $b$ quark forward-backward asymmetry with the thrust axis definition.
Unlike conventional scale setting, where the scale is simply fixed at $\mu_r=M_Z$, the PMC scale is determined by absorbing the $\beta_0$ term of the pQCD series. The resulting optimal PMC scale
is an order of magnitude smaller than the conventional choice $\mu_r=M_Z$, reflecting the small virtuality of the QCD dynamics for the process in question.
In addition, as shown by Fig.(\ref{convPMC1}), if one sets a very small scale in the conventional scale setting, the conventional result will become close to the PMC prediction, and the conventional pQCD convergence is improved compared to the case of $\mu_r=M_Z$.
Therefore the effective momentum flow for the $b$ quark forward-backward asymmetry should be $\mu_r\ll M_Z$ other than the conventionally suggested $\mu_r=M_Z$.

It is appropriate here to give a short account of the uncertainty in the resulting PMC scale determined for $A_{\rm FB}$, which is associated with the numerical uncertainties in the fixed-order NNLO inputs.
Our pure NNLO numbers for conventional scale settings listed in Table \ref{tab1} agree with those of Ref.~\cite{Bernreuther:2016ccf} at about 2\% level, compatible with the estimated Monte Carlo error of our computational setup, which merely leads to about 0.3-permile difference for $A_{\rm FB}$. After applying PMC scale-setting, this 2\% uncertainty in these pure NNLO numbers will leads to about 5\% uncertainty for the PMC scales. Since the asymmetry $A_{\rm FB}$ is completely dominated by the LO results whose numerical uncertainty is negligible, the numerical uncertainties for the PMC-improved QCD-correction factors and hence the $A_{\rm FB}$ are only about 0.6 permile.

In fact, for the $b$ quark production in electron-positron collisions at the $Z^{0}$ peak, there are two physical scales: the center-of-mass energy $\sqrt{s}=M_Z$ and the $b$ quark mass $m_b$.
The simple choice of the renormalization scale $\mu_r=M_Z$, but not the $\mu_r=m_b$, does not have a clear justification. Based on our investigation using the PMC, the effective momentum flow for the $b$ quark forward-backward should be $\mu_r\ll M_Z$.
Such a point has been noted also in other similar dynamic processes~\cite{Wang:2013vn, Sun:2018rgx}: the charmonium production in electron-positron collisions at the B factories where the effective momentum flow is shown to be around $\mu_r\sim2$ GeV, far lower than the center-of-mass energy $\sqrt{s}=10.6$ GeV. There are some additional examples to show that the conventional choice of the scale is more of a guess work. Since the renormalization scale is fixed as $\mu_r=\sqrt{s}=M_Z$ based on the conventional analysis for the event shape observables in electron-positron collisions, only one value of $\alpha_s$ at the scale $M_Z$ can be extracted~\cite{Tanabashi:2018oca}. The comprehensive analyses also show that the best choice of the effective renormalization scale should be $\mu_r\ll \sqrt{s}$~\cite{Wang:2019isi}.

Based on the massless QCD results on $A_{\rm {FB}}$ at NNLO given in Ref.~\cite{Catani:1999nf}, we have determined the PMC scale associated with the thrust axis definition to be $7.7$ GeV, which is also an order of magnitude smaller than the conventional choice $\mu_r=M_Z$. The ratio $R=|(A_{\rm FB}/A^{\rm LO}_{\rm FB})|_{\rm NNLO}/(A_{\rm FB}/A^{\rm LO}_{\rm FB})|_{\rm NLO}|$ using conventional scale-setting is $28\%$ for $\mu_r=M_Z$ and it is improved to be $12\%$ for a wide range of scale $\mu_r$ after applying PMC scale-setting. Thus, the convergence of the pQCD series for massless QCD corrections for $A_{\rm{FB}}$ is also greatly improved.

\subsection{The $b$ quark forward-backward asymmetry with quark axis definition }

\begin{table}
\begin{tabular}{|c||c|c|c|c|c|c|}
\hline
~~~ ~~~  &~~$\mu_r$~~ &~~LO~~  &~~NLO~~  &~~NNLO~~& ~~Total~~ &~~R~~ \\
\hline
~~ ~~ & $M_Z/2$ & 1 & -0.0325 & -0.0120 & 0.9556 & ~37\%~  \\
~Conv.~& $M_Z$ & 1 & -0.0291 & -0.0125 & 0.9584 & ~43\%~  \\
~~ ~~ & $2M_Z$ & 1 & -0.0263 & -0.0126 & 0.9610 & ~48\%~  \\
\hline
~PMC~ & & 1 & -0.0450 & -0.0066 & 0.9484 & ~15\%~ \\
\hline
\end{tabular}
\caption{The QCD correction factors $A_{\rm FB}/A^{\rm LO}_{\rm FB}$ for the $b$ quark forward-backward asymmetry with the quark axis definition at LO, NLO, NNLO
using the conventional (Conv.) and PMC scale settings.
The pure NNLO numbers for conventional scale settings agree with those given in Table 1 of Ref.~\cite{Bernreuther:2016ccf} at the percent level.
 \label{tab2} }
\end{table}

We now move on to discuss applying the PMC method to the calculation of $A_{\rm FB}$ with the quark axis definition where similar observations are made.
We present the QCD correction factors $A_{\rm FB}/A^{\rm LO}_{\rm FB}$ for the $b$ quark forward-backward
asymmetry with the quark axis definition at LO, NLO, NNLO using the conventional and PMC scale settings in Table \ref{tab2}.
We can see from Table \ref{tab2} that in the case of conventional scale setting, the NLO term is $-0.0291$ for $\mu_r=M_Z$ and its scale uncertainty is $21\%$ by varying $\mu_r\in[M_Z/2,2M_Z]$. The NNLO term is $-0.0125$
for $\mu_r=M_Z$ and its scale uncertainty is $5\%$ by varying $\mu_r\in[M_Z/2,2M_Z]$. The total QCD correction factor is $0.9584$ for $\mu_r=M_Z$, and its scale uncertainty is very
small as in the case of using the thrust axis definition. In addition, Table \ref{tab2} shows that the conventional predictions encounter also a slow pQCD convergence, i.e.,
the ratios $R=(A_{\rm FB}/A^{\rm LO}_{\rm FB})|_{\rm NNLO}/(A_{\rm FB}/A^{\rm LO}_{\rm FB})|_{\rm NLO}$ are very large and are changed from $37\%$ to $48\%$ by varying the scale $\mu_r\in[M_Z/2,2M_Z]$.
After applying PMC scale setting, the contribution from the NLO correction term is largely increased and the magnitude
of the NNLO correction term becomes very small due to the absorption of the $\beta_0$ term. The convergence of the PMC perturbative series is thus largely improved and the corresponding ratio is $R=(A_{\rm FB}/A^{\rm LO}_{\rm FB})|_{\rm NNLO}/(A_{\rm FB}/A^{\rm LO}_{\rm FB})|_{\rm NLO}=15\%$ for $\mu_r=M_Z$.

\begin{figure}
\centering
\includegraphics[width=0.45\textwidth]{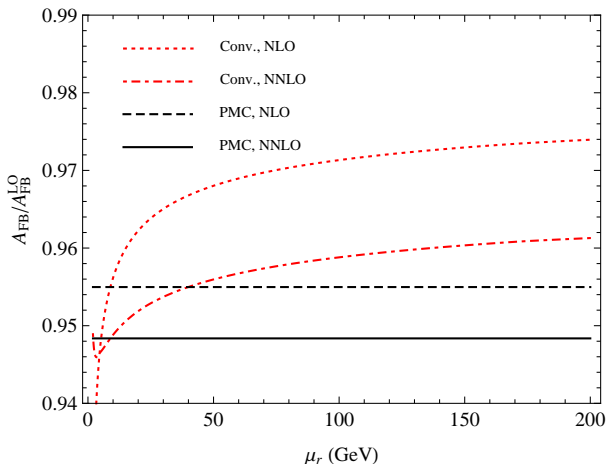}
\caption{The QCD correction factors $A_{\rm FB}/A^{\rm LO}_{\rm FB}$ versus the renormalization scale $\mu_r$ for the $b$ quark forward-backward asymmetry with the quark axis definition using the conventional (Conv.) and PMC scale settings, where the dashed and solid lines
stand for the PMC predictions up to the NLO and NNLO level, respectively; the dotted and dot-dashed lines are the corresponding results for conventional scale settings. The renormalization scale $\mu_r$ is changed in a wide range of $\mu_r\in$[2~GeV, 200~GeV].}
\label{convPMC2}
\end{figure}

In Fig.(\ref{convPMC2}) we plot the QCD correction factors $A_{\rm FB}/A^{\rm LO}_{\rm FB}$ versus the renormalization scale $\mu_r$ for the $b$ quark forward-backward asymmetry with the quark axis definition using the conventional and PMC scale settings.
It shows that using conventional scale setting, the QCD correction at NLO depends heavily on the choice of the scale $\mu_r$;
the scale dependence is decreased by the inclusion of the NNLO QCD correction, same as shown in the case of using the thrust axis definition. After using the PMC, the renormalization scale uncertainty
is eliminated and the total QCD correction is almost fixed to be $A_{\rm FB}/A^{\rm LO}_{\rm FB}=0.9484$ for a wide range of the scale $\mu_r$.

We find that the PMC scale is almost fixed to be
\begin{eqnarray}
\mu^{\rm PMC}_r&=&8.9 ~\rm {GeV}
\end{eqnarray}
for the $b$ quark forward-backward asymmetry with the quark axis definition.
Just as in the case of using the thrust axis definition, this optimal PMC scale is an order of magnitude smaller than the conventional choice of $\mu_r=M_Z$, reflecting the small virtuality of the QCD dynamics.
This shows again that the effective momentum flow for the $b$ quark forward-backward asymmetry should be $\mu_r\ll M_Z$ other than the conventionally suggested $\mu_r=M_Z$.
~\\

In the following, we show how the PMC scale setting affects the so-called bare $b$ quark forward-backward asymmetry extracted from experimental measurements.
The $A_{\rm FB}$ with QCD corrections calculated above cannot be compared directly with experimental measurements. A systematic discussion of the large deviation in $A_{\rm{FB}}$ with a full account of related  experimental issues (e.g.~those in Ref.~\cite{TheLEP/SLDHeavyFlavourWorkingGroup:2001cyl,Abbaneo:1998xt}) is beyond the scope of this paper. We then proceed with the analysis following the same procedure as adopted in Ref.~\cite{Bernreuther:2016ccf}.
We compare our QCD corrections obtained using PMC scale setting with those that were taken into account in~\cite{ALEPH:2005ab, ALEPH:2010aa, TheLEP/SLDHeavyFlavourWorkingGroup:2001cyl}. A pseudo-observable, the bare $b$-quark asymmetry $A^{0,b}_{\rm FB}$,
is determined from the measured asymmetry $A^{b,T}_{\rm FB,exp}$~\cite{TheLEP/SLDHeavyFlavourWorkingGroup:2001cyl, Abbaneo:1998xt}. In short, the asymmetry $A^{b,T}_{\rm FB,exp}$ was first corrected for QCD effects as follows
\begin{eqnarray}
A^{b,T}_{\rm FB,exp}&=&(A^{0,b}_{\rm FB})_{\rm exp}\left[1+\delta^{(1)}_Ta_s+\delta^{(2)}_Ta_s^2\right],
\label{EQ:AFBexp}
\end{eqnarray}
and then the QCD corrected ``experimental" asymmetry $(A^{0,b}_{\rm FB})_{\rm exp}$ was further corrected for
higher order electroweak corrections etc, before a value of the bare asymmetry $A^{0,b}_{\rm FB}$ was deduced.
In this way, the experimental value determined for this pseudo-observable reads as
$A^{0,b}_{\rm FB}=0.0992\pm0.0016$
and shows a large deviation, about $2.9\sigma$, from the SM fit $A^{0,b}_{\rm FB}=0.1038$~\cite{ALEPH:2005ab, ALEPH:2010aa} if the NNLO QCD corrections for massless $b$ quarks are used.
It was noted that under the thrust axis definition, the QCD correction factor
defined in Eq (\ref{EQ:AFBexp}) using the conventional scale setting
is $(1+\delta^{(1)}_Ta_s+\delta^{(2)}_Ta_s^2)=0.9646\pm0.0063$~\cite{TheLEP/SLDHeavyFlavourWorkingGroup:2001cyl},
where the error includes estimates of hadronization effects. After applying PMC scale setting, the QCD correction factor with the thrust axis definition as given in Table \ref{tab1} is $(1+\delta^{(1)}_Ta_s+\delta^{(2)}_Ta_s^2)=0.9529$, which is smaller than $0.9646$ by $1.2\%$. Our QCD correction factor with refinements from the PMC method changes $A^{0,b}_{\rm FB}=0.0992\pm0.0016$ to
\begin{eqnarray}
A^{0,b}_{\rm FB}=0.1004\pm0.0016,
\end{eqnarray}
which shows a $2.1\sigma$ deviation from the SM fit $A^{0,b}_{\rm FB}=0.1038$. The large discrepancy for $A^{0,b}_{\rm FB}$ is reduced but only slightly to $2.6\sigma$ when full $b$ quark mass effects are included~\cite{Bernreuther:2016ccf}. We thus conclude that the PMC-improved calculation diminishes the well known tension between the bare $b$ quark forward-backward asymmetry and its SM fit from 2.6$\sigma$ to 2.1$\sigma$. It is noted that a new fit result $A^{0,b}_{\rm FB}=0.1032\pm0.0004$ is given in Ref.~\cite{Baak:2014ora}, our PMC-improved result show a $1.8\sigma$ deviation from this new SM fit.

\section{Summary }
\label{sec:4}

In the case of conventional scale setting, a fixed-order pQCD result contains an inherent renormalization scheme-and-scale dependence. In contrast, the PMC procedure provides a rigorous method for unambiguously setting the renormalization scale, and the resulting PMC scale reflects the effective virtuality of the underlying QCD subprocesses. The PMC results are independent of the choice of the initial renormalization scale and the choice of renormalization scheme.

In this paper, we have employed the PMC method to fix the $\alpha_s$-running behavior of the QCD corrections at NNLO to the $b$ quark forward-backward asymmetry $A_{\rm{FB}}$.
The conventional results based on conventional scale setting are plagued by the scale $\mu_r$ uncertainty and show a slow pQCD convergence due to the presence of the renormalon terms.
The previous results for the pseudo-observable $A^{0,b}_{\rm FB}$ show a large discrepancy of $2.9$ standard deviations between the experimental determination and the corresponding SM prediction, which is reduced but only slightly to $2.6\sigma$ when the NNLO QCD corrections are included with full $b$ quark mass effects using conventional scale setting~\cite{Bernreuther:2016ccf}.
After applying the PMC procedure, the so-refined pQCD prediction eliminates the renormalization scale uncertainty. The convergence of the pQCD series is also greatly improved, since the renormalon divergences are eliminated as well.
We have determined the PMC scale for $A^{0,b}_{\rm FB}$ at the $Z^0$ peak is $\sim 9$ GeV, an order of magnitude smaller than the conventional choice $\mu_r=M_Z$, reflecting the small virtuality of the QCD dynamics for this observable. The PMC-improved result for the bare $b$ quark forward-backward asymmetry reads as $A^{0,b}_{\rm FB}=0.1004\pm0.0016$, which diminishes the well known tension between its experimental determination and the respective SM fit to $2.1\sigma$. Thus we think it is important to apply a proper scale-setting approach, such as PMC, to the pQCD series so as to achieve a more precise SM prediction, or a better constraint on the possible new physics.

\hspace{1cm}

{\bf Acknowledgements}: This work was supported in part by the Natural Science Foundation of China under Grants No.11625520, No.11705033, No.11905056 and No.11947406; by the Project of Guizhou Provincial Department under Grant No.KY[2017]067.

\end{document}